%% file: main.tex
\documentclass[runningheads]{llncs}
\usepackage{colortbl}

\usepackage{xcolor}
\usepackage{hyperref}
\usepackage{array,multirow,color, graphicx, makecell, arydshln, multirow, makecell}

\usepackage{graphicx}

\makeatletter
\newcommand{\printfnsymbol}[1]{%
  \textsuperscript{\@fnsymbol{#1}}%
}
\makeatother

\begin{document}
\title{Feather-Light Fourier Domain Adaptation in Magnetic Resonance Imaging}
\titlerunning{Feather-Light Fourier Domain Adaptation in MRI}
\author{
    Ivan Zakazov \thanks{Equal contribution.} \inst{1, 2} \and
    Vladimir Shaposhnikov \printfnsymbol{1} \inst{1, 2} \and
    Iaroslav Bespalov \inst{1, 2} \and
    Dmitry V. Dylov \thanks{Corresponding author.}\inst{2}
}

\authorrunning{I. Zakazov, V. Shaposhnikov et al}

\institute{
    Philips Research, Moscow, Russia
    \and
    Skolkovo Institute of Science and Technology, Moscow, Russia
    \\
    \email{D.Dylov@skoltech.ru}
}

\maketitle              
\begin{abstract}
Generalizability of deep learning models may be severely affected by the difference in the distributions of the train (\textit{source domain}) and the test (\textit{target domain}) sets, \textit{e.g.}, when the sets are produced by different hardware. 
As a consequence of this \textit{domain shift}, a certain model might perform well on data from one clinic, and then fail when deployed in another. We propose a very light and transparent approach to perform \textit{test-time domain adaptation}. 
The idea is to substitute the \textit{target} low-frequency Fourier space components that are deemed to reflect the style of an image. 
To maximize the performance, we implement the \textit{``optimal style donor''} selection technique, and use a number of \textit{source} data points for altering a single \textit{target} scan appearance (\textit{Multi-Source Transferring}). We study the effect of severity of domain shift on the performance of the method, and show that our \textit{training-free} approach reaches the state-of-the-art level of complicated deep domain adaptation models. The code for our experiments is released\footnote{https://github.com/kechua/Feather-Light-Fourier-Domain-Adaptation/}.

\end{abstract}
\section{Introduction}
\input{Sections/intro}
\label{sec:intro}

\section{Related Work}
\input{Sections/relatedwork}

\label{sec:related}

\section{Method}
\input{Sections/method}

\label{sec:method}

\section{Experiments}

\input{Sections/experiments}

\label{sec:experiments}

\section{Conclusion}
\input{Sections/conclusion}
\label{sec:conclusion}

\section*{Acknowledgements}
\input{Sections/thanks}
\label{sec:thanks}

\bibliographystyle{ref_style}
\bibliography{main}
\end{document}


\title{Supplementary material for "Feather-Light Fourier Domain Adaptation in Magnetic Resonance Imaging"}

\titlerunning{FDA in MRI}

\author{}
\institute{}
\maketitle

\begin{figure*}[h!]
\begin{center}
\includegraphics[width=\textwidth]{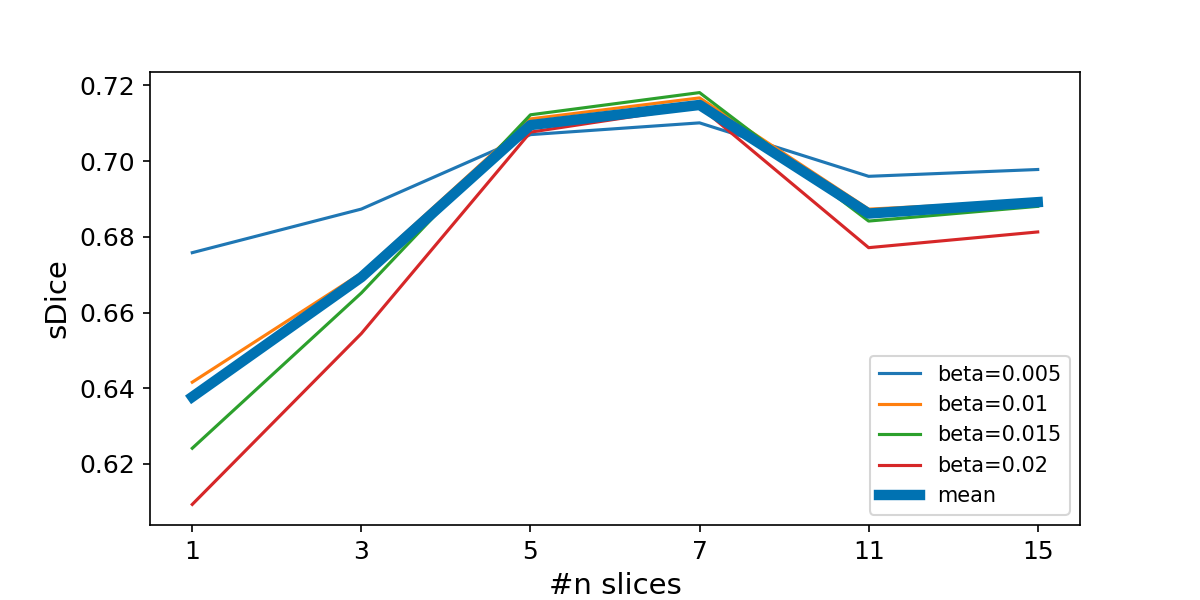}
\caption{Dependence of the Multi-Source Transferring effectiveness on the number of the source slices. We set $n=7$ as it proves to be the optimal value}
\label{fig:da_comparison}
\end{center}
\end{figure*}

\begin{figure*}[h!]
\begin{center}
\includegraphics[width=\textwidth]{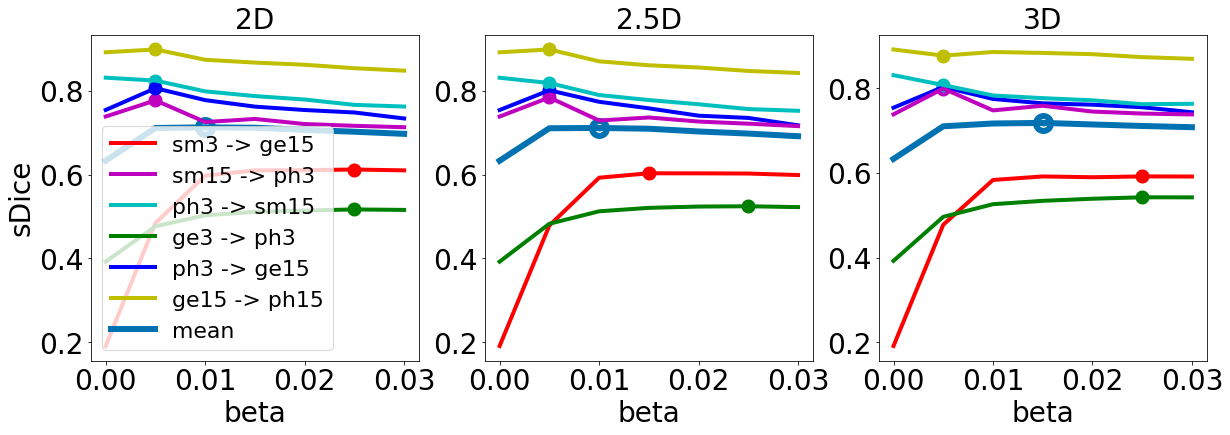}
\caption{Choosing the optimal $\beta$ value on validation set in both \textbf{Averaged Optimal} and \textbf{Optimal per Pair} scenarios. In case $\beta=0$ is the global maxima, we set $\beta=0.005$}
\label{fig:da_comparison}
\end{center}
\end{figure*}

\newpage

\begin{figure*}[h!]
\begin{center}
\includegraphics[width=\textwidth]{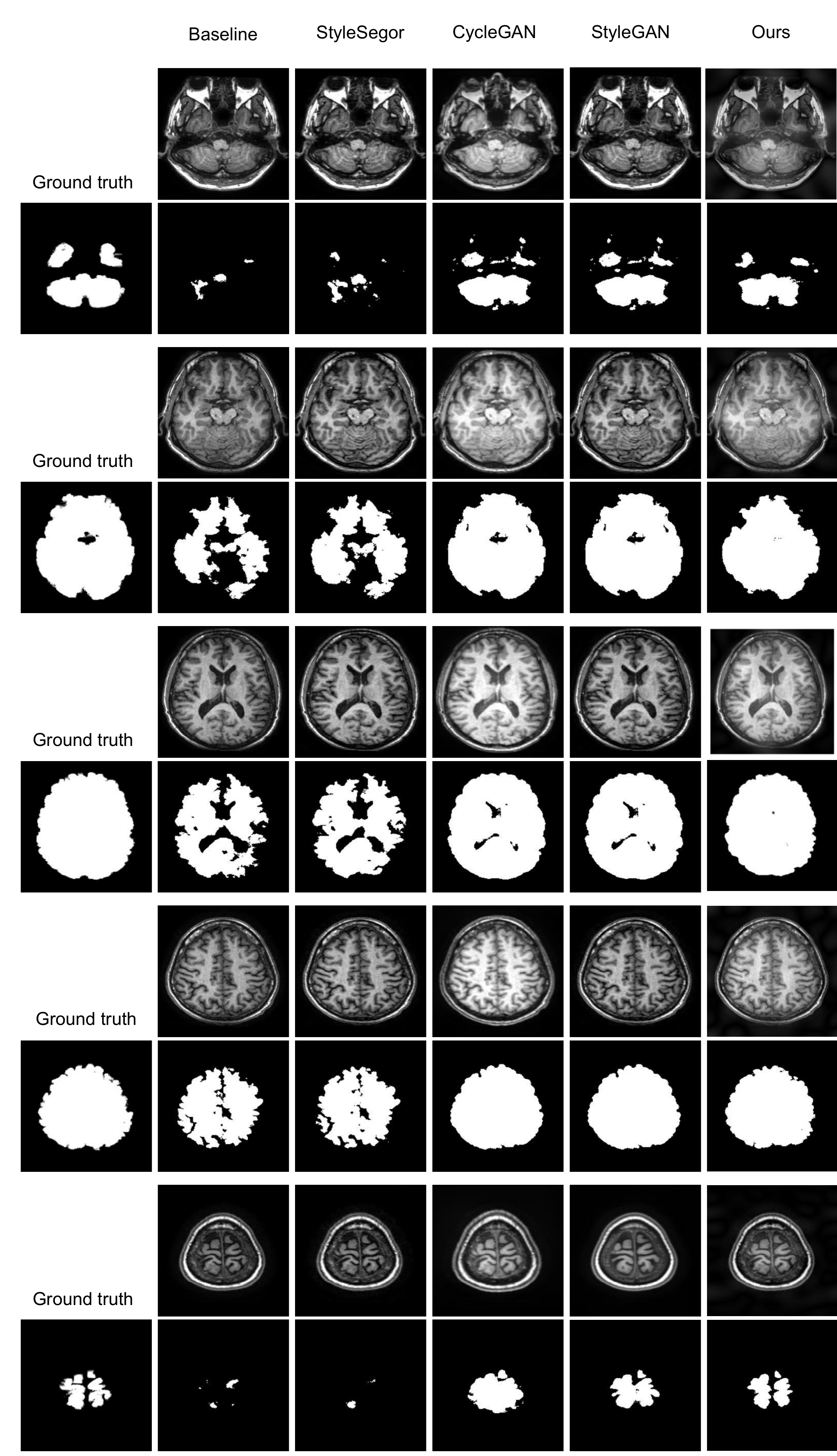}
\caption{Visual comparison of various approaches (in our method, $\beta$=0.03), tackling strong domain shift. Different slices of a single scan are considered}
\label{fig:da_comparison}
\end{center}
\end{figure*}

%% file: Sections/intro.tex
Magnetic Resonance Imaging (MRI) has become an irreplaceable tool in healthcare thanks to its capacity to produce high-resolution scans without ionizing radiation. The widespread use of the modality has helped to accumulate large volumes of miscellaneous imaging data, which have been fueling development of machine- and deep-learning methods, aiming to mimic diagnostic decisions. However, the real-life deployment of these methods is often hindered by an issue known as \textit{Domain Shift}, which originates from a possible difference in train (\textit{source}) and test (\textit{target}) distributions. This difference might occur whenever \textit{source} and \textit{target} datasets are acquired with different machines or research protocols, and entails a need for proper \textit{Domain Adaptation} (DA) \cite{Guan}. 

To this end, modern DA in medical imaging includes a plethora of shallow and deep models \cite{Guan}. Existing shallow DA methods are somewhat rudimentary, requiring human-engineered features \cite{shallow1,shallow2}, and generally lagging in performance, while the deep ones are often heavy, slow, and barely interpretable (albeit accurate) \cite{Kamnitsas2017UnsupervisedDA,perone2019unsupervised,Wollmann}.

One trait shared by all of the aforementioned methods is that they operate in the \textit{image space}. It is only recently that the community has started to realize the potential of operating in \textit{k}-space (also referred to as \textit{Fourier space} or \textit{spectrum} of an image) for tackling domain shift in MRI data with \cite{FedDG,CardicFDA} being the only two studies we were able to find. This is surprising given that MRI is a modality that yields \textit{k}-space data representation by design.

In \textit{k}-space, specific spectral components are responsible for different properties of an image, \textit{e.g.}, the high frequency components accentuate the edges and enhance details, while the low ones control contrast and the large-scale content. Besides, it is known that the semantic information is mostly stored in the phase component of the spectrum.
This suggests an efficient strategy for tackling DA via ``mixing'' the \textit{source} and the \textit{target} spectra. 
The purpose of this "mixing" (\textit{Fourier Domain Adaptation} or \textit{FDA}) is to transfer the style while \textbf{preserving the patient-specific content}, thereby compensating for the DA shift.

The idea is borrowed from the natural image domain \cite{fda_main} and adapted to MRI volumes, entailing a new \textit{``optimal style donor''} selection module and a \textit{multi-source} transferring routine (we use multiple \textit{source} images when transferring the style to a given \textit{target} image to further improve the performance).

We call our method \textit{feather-light} because, unlike the modern go-to approaches, it \textbf{does not involve any training}, as we simply transfer the \textit{k}-space components, characterizing the style of the \textit{source} domain, to a \textit{target} scan during the test time. 
Despite its simplicity, the method performs on par with complicated deep DA models, such as those based on Generative Adversarial Networks (GANs) \cite{welander2018generative}.
Notably, the proposed method is also \textit{interpretable} as it directly shows which style-carrying source frequencies alleviate the domain shift.

%% file: Sections/relatedwork.tex
A large part of Deep Domain Adaptation methods could be split into feature-level \cite{ganin2016domainadversarial,ma2019neural} and image-level approaches. Among the image-level ones, the majority exploit the idea of GANs \cite{CardicFDA,joshi2021ai,welander2018generative} for eradicating the difference in distribution between images from various domains. GANs, however, are difficult to train, lack explainability and might produce undesirable artefacts, which pose an even greater problem in the medical imaging context

Fourier Domain Adaptation (\textbf{FDA}) \cite{fda_main} provides a feasible alternative to GANs, as image-to-image translation, performed via low-frequency spectra components swap (amplitudes only), is simple, predictable and yet yields SOTA-level results on the natural images. This method has been adapted for the medical imaging, with the earliest application being mitigation of domain shift, appearing in the synthetic ultrasound images \cite{ultrasoundFDA}. In \cite{CardicFDA} the authors applied FDA-based augmentation technique for the cardiac MRI segmentation, with the novelty being swapping both amplitudes and phases, which apparently is not very stable and may lead to changes in the image semantics \cite{yang2020phase}. \cite{FedDG} applies FDA to federated learning in order to generate images exhibiting distribution characteristic of other clients, while \cite{feng2021fiba} uses FDA as a proof-of-concept tool for obtaining "poisoned" images, challenging for neural networks.    

In \cite{LIU2021102052} the authors solve automatic polyp detection task via combining feature-level adaptation with FDA, while further improving the FDA component with sampling "matching" source and target image pairs. The closer a target image is to the source one in terms of cosine similarity of their deep ResNet-50 features, the greater is the "match" probability. We note that adding an additional deep model to the pipeline makes it more complicated, while we strive for simplicity. 

One dismissed idea in the FDA area is the one we propose to denote \emph{multi-source transfer}, \textit{i.e.}, performing a number of \textit{k}-space components swaps with a single \emph{target} image and multiple \emph{source} images followed by averaging of the down-stream task predictions for various versions of the changed \emph{target}.

%% file: Sections/method.tex
\begin{figure}[t]
\centering
\includegraphics[width=\textwidth]{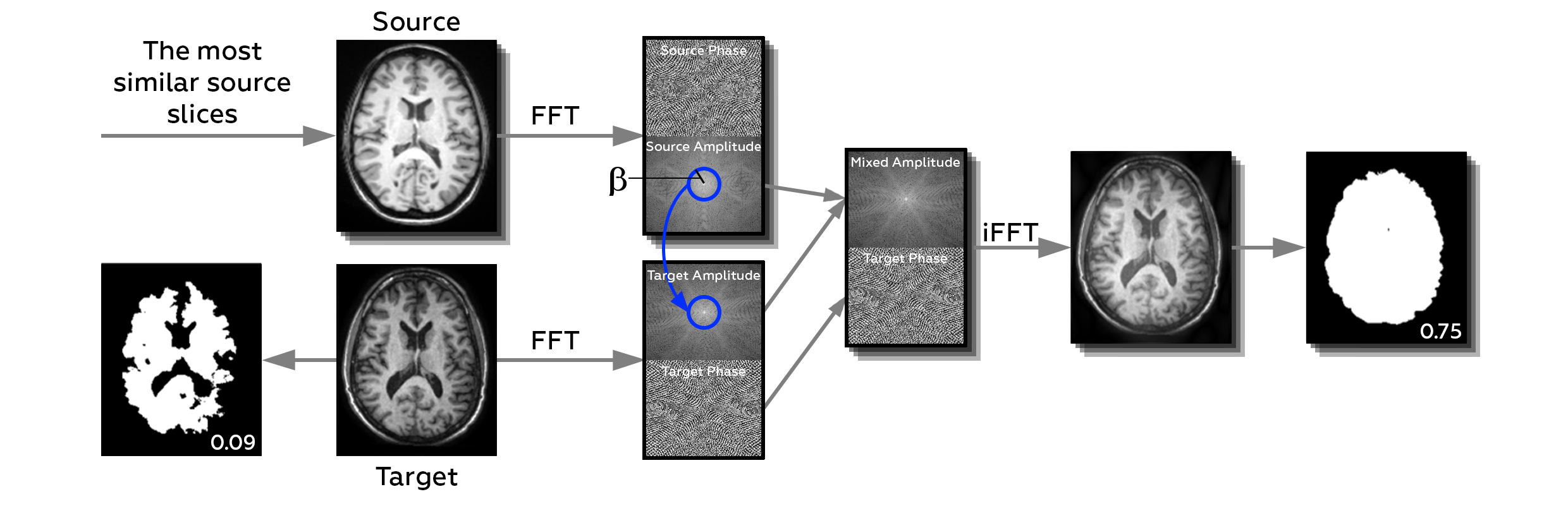}
\caption{Fourier Domain Adaptation (FDA) for the Brain Segmentation task.}
\label{fig:fda}
\end{figure}

We base our approach on the Fourier Domain Adaptation technique \cite{fda_main} and summarize it in Fig. \ref{fig:fda}. This method consists of swapping the low-frequency amplitudes of an image spectrum with those of another image, the style of which should be borrowed. As amplitudes of the low-frequency spectrum components are mostly related to the low-level image characteristics, defining the style, this procedure is expected to align the \emph{source} and the \emph{target} distributions, thus compensating for the \emph{domain shift} between them.

While in \cite{fda_main}, the \emph{source} data is transferred to the \emph{target} style, and the deep neural network is then trained on the $D^{s \rightarrow t}$ dataset, we note that in the clinical setting it would mean re-training the model for each new $target$ domain (e.g., a new clinic), which might complicate certification and clinical deployment. Moreover, the $target$ data (\textit{e.g.}, data from a hospital we need to adjust the model for) may appear to be scarce, which limits capabilities of the self-supervised training on \emph{target}, another component of the original method.

Instead of $D^{s \rightarrow t}$, we focus on the $D^{t \rightarrow s}$ adaptation with a single \emph{source model} used for various \emph{targets}, which are transferred into the \emph{source} style during the test time. Mathematically speaking, we carry out the following procedure: $t_{new} = f_{FDA}(s_{i}, t) = 
\mathcal{F}^{-1}\left(\left[M_{\beta} \circ \mathcal{F}^{A}\left(s_{i}\right)+\left(1-M_{\beta}\right) \circ \mathcal{F}^{A}\left(t\right), \mathcal{F}^{P}\left(t\right)\right]\right)$.

The phase component of a spectrum remains intact, while the source style is injected with the low-frequency amplitudes we "cut out" with $M_{\beta}$ (Fig.\ref{fig:fda}).  

As there is no additional training required, this setting is much more light-weight than the original one, but it is also more challenging in terms of reaching the optimal performance. To this end, we improve the method in the following ways:

\let\labelitemi\labelitemii
\begin{itemize}
    \item We propose to carry out a multitude of $t \rightarrow s_{i}$ swaps (Multi-Source Transfer) with the final result for slice t calculated as $\sum_{i=1}^{n_{MST}} net(f_{FDA}(s_{i}, t)) / n_{MST} $, where $f_{FDA}(s_{i}, t)$ is the FDA procedure, performed on $s_{i}$ and $t$; $n_{MST}$ reflects the number of \textit{source} slices used for the style transfer (after preliminary experiments we set $n_{MST}=7$)
    \item We design an approach for picking the optimal \emph{source} slices  
\end{itemize}

\begin{figure}[h!]
\centering
\includegraphics[width=\textwidth]{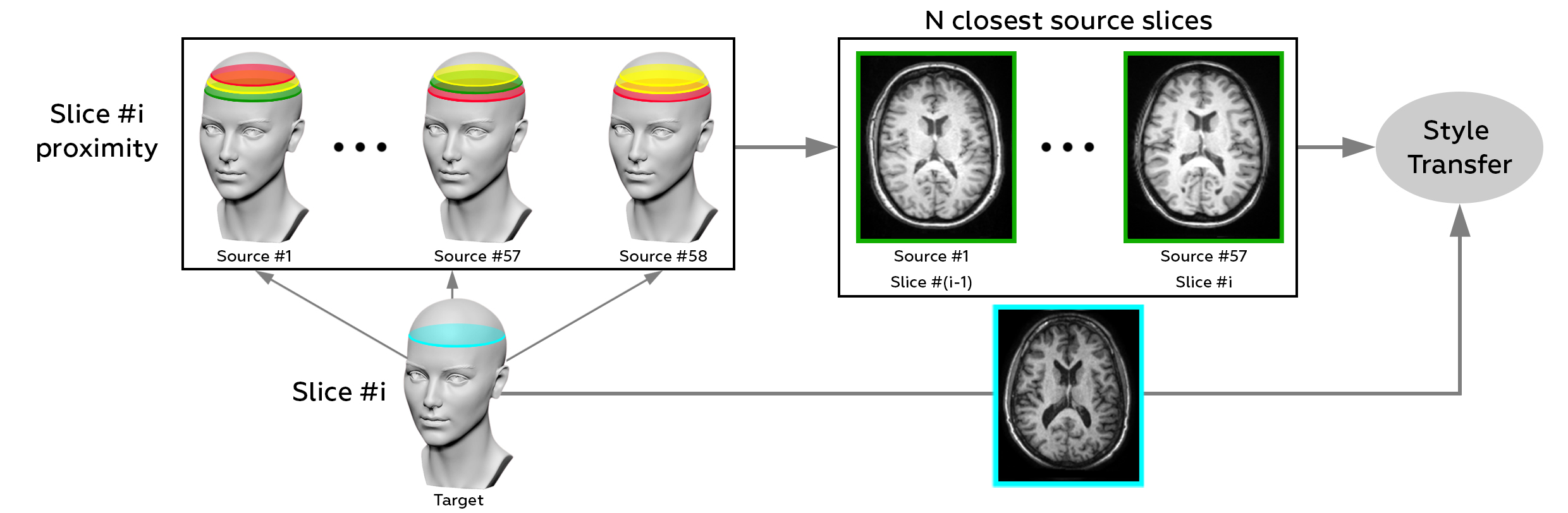}
\caption{Multi-Source Transfer (MST) + SR-SIM source choice in the \textbf{2.5D} fashion.}
\label{fig:pick_source}
\end{figure}

The intuition behind the latter feature is that as swapping the spectral components inevitably leads to artefacts, we should minimize this detrimental effect by choosing \emph{source} and \emph{target} which are as close as possible in terms of their semantics. To do so, we assess the "closeness" with the Spectrual Residual Similarity (SR-SIM) semantic similarity measure \cite{SRSIM} (Fig.\ref{fig:pick_source}). 

We consider several approaches to the optimal \emph{source "style donor"} search for the slices, belonging to the \emph{target} scan. Firstly, we may average the similarity score between the corresponding slices in $2$ scans, thus obtaining the scan-to-scan similarity score ($sim(s,t) = \sum_{i=1}^{n_{slices}} sim(s_{i}, t_{i}) / n_{slices} $). We then select for Domain Adaptation $n_{MST}$ most similar \emph{source} scans (\textbf{3D} similarity). 

We note that choosing the "style donors" on the scan level might introduce unnecessary constraint. Alternatively,  \textit{source} "style donors" for various $t_i$ slices of the target scan $t$ may come from different \textit{source} scans. In this case, for $i \in (1,  n_{slices})$ we look for the slices closest to $t_i$ among $(s_i^1, s_i^2, ... , s_i^{n_{scans}})$ (\textbf{2D}). A natural extension to this approach is broadening this set to $(s_{i-m}^1, s_{i-m+1}^1, ... , s_{i}^1,$ $s_{i+1}^1, s_{i+m}^1, s_{i-m}^2, ... , s_{i+m}^{n_{scans}})$, which we refer to as \textbf{2.5D} (Fig.\ref{fig:pick_source}). We set $m=2$.

%% file: Sections/experiments.tex
\subsection{Technical details}
\label{ssec:exp:tech}

We conduct all the experiments on a public brain MR dataset called CC359 \cite{souza2018open}, which is formed of $359$ scans and various masks, among which are the \emph{brain segmentation} masks. The scans are produced by one of $6$ MRI machines (Siemens, Philips, GE; 1.5T or 3T each), and thus fall into one of 6 domains of approximately equal size ($60$ or $59$ scans). We perform affine registration of all the scans to MNI152 template using the FSL software \cite{FSL2,FSL1}, and subsequently normalize voxel intensities to $[0,1]$. We use the \emph{Surface Dice Score} \cite{nikolov2018deep} as it appears to be a more reliable indicator of the brain segmentation quality than the standard Dice Score \cite{shirokikh2020first}.

We solve the brain segmentation task with 2D U-Net with residual blocks, which we train for $100$ epochs ($100$ iterations per epoch), using SGD optimizer with Nesterov momentum of $0.9$, combination of BCE and dice losses (weighted with $0.4$ and $0.6$ coefficients), and learning rate of $10^{-3}$, reduced to $10^{-4}$ at epoch $80$. We train the networks on $(256, 256)$ crops grouped in batches of $16$ samples. The crops are sampled randomly at each iteration.   

\input{Tables/Domain_table}

We follow the methodology of the original Fourier Domain Adaptation (FDA) paper \cite{fda_main} with respect to the k-space swapping technique, with the notable difference of using the circular crop instead of the rectangular one, which is to take into account the radial symmetry of the spectrum components amplitudes.

\subsection{Naive model transfer}
\label{ssec:exp:naive}

Firstly, we consider a simple case of \emph{no Domain Adaptation} applied. In this regard, we train $6$ \emph{base models} on the corresponding domains, designating all but $2$ \emph{source} scans for training (these $2$ are to ensure reaching the loss plateau when training). We then transfer these source-trained models to \emph{unseen domains}, thus considering $30$ source-target pairs. We calculate each transferred model performance on the \emph{target test set} of $10$ images. Besides, we also use $3$-fold cross-validation to assess the model performance on the domain it was trained on.

As may be seen from Fig. \ref{tab:baseline}, the magnitude of \emph{Domain Shift}, \textit{i.e.}, the performance variability between the transferred model and the one which was initially trained on some domain changes significantly between the source-target pairs. As conducting subsequent experiments on all $30$ source-target pairs is computationally prohibitive, we decide to concentrate on 3 clusters, representing \textbf{severe} domain shift, \textbf{medium} domain shift and the \textbf{subtle} one. We sort the source-target pairs by the metric decline magnitude, and pick 2 pairs per cluster from the top, bottom, and middle of this sorted list.   

\input{Tables/baselines_table}

\subsection{Choosing the optimal $\beta$}
\label{ssec:exp:beta}

One of the most important FDA design choices is choosing the size of the swapping window $\beta$. Specifically for this purpose we designate another $10$ \emph{target} scans per source-target pair, on which the grid search over various $\beta$ values is performed. We consider $2$ strategies of devising the optimal $\beta$, which correspond to $2$ actual clinical set-ups:

\let\labelitemi\labelitemii
\begin{itemize}
    \item \textbf{Optimal per Pair}. Picking the $\beta$, which proved to be the optimal one for each source-target pair. This set-up is motivated by the scenario, in which at least some target domain data (\textit{e. g.}, data coming from a new clinical center) is available and labelled, and thus may be used for setting the optimal $\beta$, peculiar to this source-target pair
    \item  \textbf{Averaged Optimal}. Picking the $\beta$, based on the grid-search results, averaged over all pairs, which corresponds to a broader scenario of setting a single "standard" beta for all the pairs    
\end{itemize}

\input{Tables/D_Approach_table}

\subsection{Results and Discussion}
\label{ssec:exp:res}

As was discussed in Section \ref{sec:method}, we consider $3$ approaches to picking the "best" source slices, which we denote \textbf{2D}, \textbf{2.5D}, and \textbf{3D}. Furthermore, in line with \ref{ssec:exp:beta} we devise $\beta$ from the \emph{target} validation set by means of either \emph{global averaging} or \emph{averaging per pair}. The corresponding results are presented in Table \ref{tab:results} in comparison with the SOTA-level baselines of CycleGAN \cite{welander2018generative}, StyleGAN \cite{CardicFDA} and Style-Segor \cite{ma2019neural}. We also consider another light-weight baseline \cite{GhiasiLKDS17}. 

We perform the ablation study (Table \ref{tab:ablation}), comparing (a) SR-SIM chosen sources + Multi-source transfer (MST) (b) Multi-source transfer (MST) (c) A "simple" swap.

\textbf{2D \textit{vs.} 2.5D \textit{vs.} 3D}. Interestingly, no significant difference could be observed between various SR-SIM-based source choice approaches. Subsequently, we concentrate on the \textbf{3D} approach, as it appears to be both more intuitive and marginally better than the others

\textbf{$\beta$: Averaged Optimal \textit{vs.} $\beta$: Optimal per Pair}. Picking $\beta$ on \textit{target} Validation set in a pair-wise fashion gives only a minor advantage over devising it from the averaged \textit{target} validation $sDice(\beta)$ curve. Besides, the latter set-up does not require adjusting $\beta$ for a particular pair on a labelled subset of target data, and thus is more relevant in the clinical practice. Therefore, from now on we concentrate on the \textit{Averaged Optimal} results analysis.

\textbf{Our method (3D; $\beta$: Averaged Optimal) \textit{vs.} Baselines}. While our method is outperformed by \textbf{CycleGAN} on \emph{severe \#2} pair and by \textbf{StyleSegor} on \emph{subtle \#1} pair, it is the only one demonstrating good performance across all the data shift magnitude range, since \textbf{GANs} fail to preserve even the "naive" swap quality (\textbf{no DA}) in case of low domain shift and \textbf{StyleSegor} is barely improving the score in case of strong domain shift. Fast artistic image stylization \cite{GhiasiLKDS17}, another light-weight method we consider as a baseline, does not demonstrate sufficiently good performance

\begin{figure*}[h]
\begin{center}
\includegraphics[width=\textwidth]{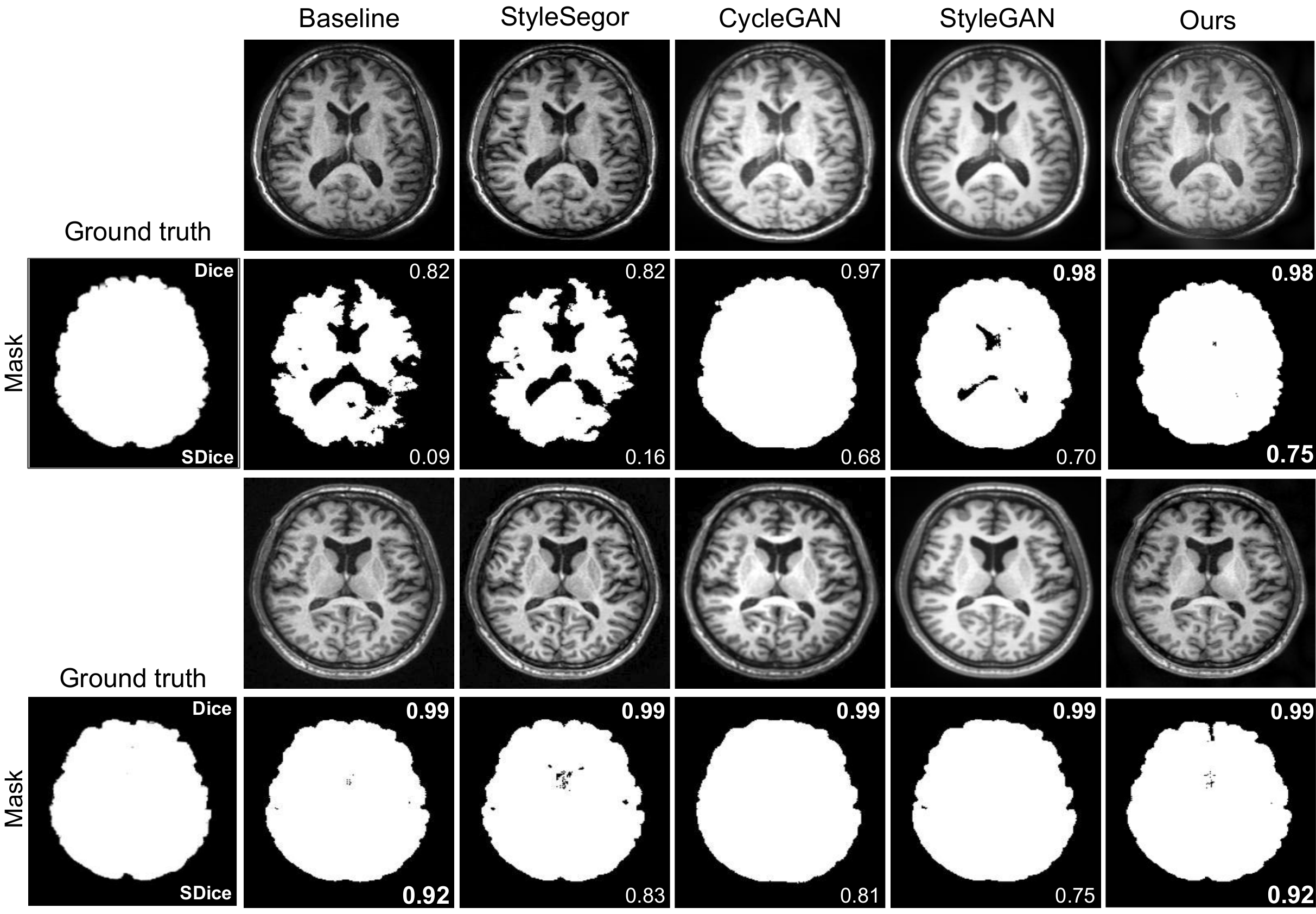}
\caption{Visual comparison of various approaches. In this particular case, we set $\beta$=0.03 }
\label{fig:da_comparison}
\end{center}
\end{figure*}

\textbf{Ablation Studies}. As could be seen in Table \ref{tab:ablation}, both introducing Multi-Source Transfer and combining it with the SR-SIM-based source choice improves the score on average with the positive effect of the "smart" source choice substantial for the instances of severe domain shift.

In Fig. \ref{fig:da_comparison}, we visually compare our approach with the baselines, considering the cases of severe (top) and subtle (bottom) domain shift. For illustration purposes, we apply the method to the middle-positioned slices. Notably, in case of severe domain shift GANs alter the appearance much more significantly, which might explain the decreasing score.

%% file: Tables/Domain_table.tex
\newcommand{\STAB}[1]{\begin{tabular}{@{}c@{}}#1\end{tabular}}
\begin{table*}[h!]
\centering
\begin{tabular}{|c|c|c|c|c|c|c|c|} 
\hline
\multicolumn{2}{|l|}{\multirow{2}{*}{}}  & \multicolumn{6}{c|}{Source domains}\\ 
\cline{3-8}
\multicolumn{2}{|l|}{} & sm15 & sm3 & ge15 & ge3 & ph15 & ph3 \\ 
\hline
\multirow{6}{*}{{\STAB{\rotatebox[origin=c]{90}{Target domains}}}} & sm15 & {\cellcolor[rgb]{0.827,0.843,0.812}}0.90  $\pm$  0.03 & 0.57 $\pm$ 0.18 & 0.83 $\pm$ 0.07 & 0.54 $\pm$ 0.18 & 0.78 $\pm$ 0.09 & \textcolor[rgb]{0,0.504,0}{0.84 $\pm$ 0.03}   \\ 
\cline{2-8}
& sm3  & 0.81 $\pm$ 0.04 & {\cellcolor[rgb]{0.827,0.843,0.812}}0.90 $\pm$ 0.02 & 0.78 $\pm$ 0.03 & 0.63 $\pm$ 0.07 & 0.80 $\pm$ 0.05 & 0.78 $\pm$ 0.03 \\ 
\cline{2-8}
& ge15 & 0.61 $\pm$ 0.17 & \textcolor{red}{0.11 $\pm$ 0.06} & {\cellcolor[rgb]{0.827,0.843,0.812}}0.90 $\pm$ 0.03  & 0.40 $\pm$ 0.16 & 0.51 $\pm$ 0.18 & \textcolor[rgb]{1,0.647,0}{0.67 $\pm$ 0.15}  \\ 
\cline{2-8}
& ge3  & 0.84 $\pm$ 0.03 & 0.44 $\pm$ 0.14 & 0.78 $\pm$ 0.07 & {\cellcolor[rgb]{0.827,0.843,0.812}}0.91 $\pm$ 0.03 & 0.76 $\pm$ 0.1 & 0.78 $\pm$ 0.03 \\ 
\cline{2-8}
& ph15 & 0.83 $\pm$ 0.06 & 0.45 $\pm$ 0.1 & \textcolor[rgb]{0,0.502,0}{0.87 $\pm$ 0.03} & 0.42 $\pm$ 0.17 & {\cellcolor[rgb]{0.827,0.843,0.812}}0.91 $\pm$ 0.03 & 0.79 $\pm$ 0.03 \\ 
\cline{2-8}
& ph3  & \textcolor[rgb]{1,0.647,0}{0.74 $\pm$ 0.12} & 0.40 $\pm$ 0.12 & 0.62 $\pm$ 0.12 & \textcolor{red}{0.39 $\pm$ 0.12} & 0.56 $\pm$ 0.12 & {\cellcolor[rgb]{0.827,0.843,0.812}}0.88 $\pm$ 0.04 \\
\hline
\end{tabular}
\caption{Naive transferring (no Domain Adaptation applied).}
\label{tab:baseline}
\end{table*}

%% file: Tables/baselines_table.tex

\newcolumntype{x}[1]{%
>{\centering\hspace{0pt}}p{#1}}%

\begin{table}[b]
\centering
\begin{tabular}{|c|c|x{0.85cm}|x{0.85cm}:x{0.85cm}:x{0.75cm}:x{0.75cm}|x{0.75cm}:x{0.75cm}:x{0.75cm}|x{0.75cm}:x{0.75cm}:x{0.75cm}|} 
\hline
\multicolumn{2}{|c|}{\multirow{2}{*}{}} &  & \multicolumn{4}{c|}{Baselines}& \multicolumn{3}{l|}{$\beta$ : aver. optimal} & \multicolumn{3}{l|}{$\beta$ : opt. per pair}                                   \tabularnewline
\multicolumn{2}{|c|}{} & No DA & StyleSegor & Cycle & Style & Fast & 3D & 2.5D & 2D & 3D & 2.5D & 2D \tabularnewline 
\hline
\multirow{6}{*}{\STAB{\rotatebox[origin=c]{90}{Shift severity}}} & severe \#1 & 0.11 & 0.11& 0.50
& 0.46 & 0.15 & 0.57 & 0.57 & 0.57 & 0.57 & 0.58 & \textbf{0.59} \tabularnewline

& \multicolumn{1}{c|}{severe \#2} & 0.39 & 0.46 & \textbf{0.64} & 0.58 & 0.12 & 0.50 & 0.48 & 0.47 & 0.51 & 0.48 & 0.48  \tabularnewline

\cline{2-13}
& \multicolumn{1}{c|}{medium \#1} & 0.67 & 0.66 & 0.70 & 0.64 & 0.15 & 0.72 & 0.73 & 0.74 & \textbf{0.77} & \textbf{0.77} & \textbf{0.77}  \tabularnewline

& \multicolumn{1}{c|}{medium \#2} & 0.74 & 0.69 & 0.69 & 0.61 & 0.11 & 0.72 & 0.69 & 0.68 & \textbf{0.75} & 0.74 & 0.73  \tabularnewline

\cline{2-13}
& \multicolumn{1}{c|}{subtle \#1} & 0.84 & \textbf{0.85} & 0.60 & 0.41 & 0.17 & 0.81 & 0.81& 0.82 & 0.83 & 0.83 & 0.84  \tabularnewline

& \multicolumn{1}{c|}{subtle \#2} & \textbf{0.87} & 0.82 & 0.46 & 0.55 & 0.11 & 0.85 & 0.83 & 0.84 & \textbf{0.87} & 0.86 & 0.86  \tabularnewline

\hline

& \multicolumn{1}{c|}{average} & 0.60 & 0.60 & 0.60 & 0.54 & 0.14 & 0.7 & 0.68 & 0.69 & \textbf{0.72} & 0.71 & 0.71  \tabularnewline
\hline
\end{tabular}
\caption{Comparison of the performance of various proposed methods with the baselines.
\textit{Style} is for StyleGAN \cite{9156570}, \textit{Cycle} is for CycleGAN \cite{welander2018generative}, \textit{Fast} is for artistic stylization network \cite{GhiasiLKDS17}.}
\label{tab:results}
\end{table}

%% file: Tables/D_Approach_table.tex
\newcolumntype{x}[1]{%
>{\centering\hspace{0pt}}p{#1}}%

\begin{table}[h!]
\centering
\begin{tabular}{|c|c||x{1.4cm}:x{1.4cm}:x{1.4cm}|x{1.4cm}:x{1.4cm}:x{1.4cm}|} 
\hline
\multicolumn{2}{|c|}{\multirow{2}{*}{}} & \multicolumn{3}{c|}{$\beta$ : averaged optimal} & \multicolumn{3}{c|}{$\beta$ : optimal per pair}                                  \tabularnewline

\multicolumn{2}{|c|}{} & SR-SIM+ & MST & None & SR-SIM+ & MST & None    \tabularnewline
\hline
\multirow{6}{*}{\STAB{\rotatebox[origin=c]{90}{Shift severity}}} & \multicolumn{1}{c|}{severe \#1} & \textbf{0.57} & 0.41 & 0.40 & \textbf{0.59} & 0.57& 0.53   \tabularnewline
& \multicolumn{1}{c|}{severe \#2}& \textbf{0.47}& 0.42 & 0.41 & \textbf{0.48} & 0.41 & 0.41  \tabularnewline
\cline{2-8}
& \multicolumn{1}{c|}{medium \#1} & 0.74 & \textbf{0.78} & 0.76 & 0.77 & \textbf{0.78} & 0.76   \tabularnewline
& \multicolumn{1}{c|}{medium \#2} & \textbf{0.68} & 0.69 & 0.65 & \textbf{0.73} & 0.69 & 0.65   \tabularnewline
\cline{2-8}
& \multicolumn{1}{c|}{subtle \#1} & 0.82 & \textbf{0.85} & 0.82 & 0.84 & \textbf{0.85} & 0.82   \tabularnewline
& \multicolumn{1}{c|}{subtle \#2}  & 0.84 & \textbf{0.85} & 0.84 & 0.86 & \textbf{0.85} & 0.84   \tabularnewline
\hline
&\multicolumn{1}{c|}{average} & \textbf{0.69} & 0.67 & 0.65 & \textbf{0.71} & 0.69 & 0.67   \tabularnewline
\hline
\end{tabular}
\caption{The ablation study.}
\label{tab:ablation}
\end{table}

%% file: Sections/conclusion.tex
We present a novel Fourier-based Domain Adaptation method, which requires neither any training, nor incorporating any additional deep components into the pipeline. We consider various domain shift severity scenarios, and show that our method performs consistently across all of them, outperforming SOTA-level GANs in case of \textit{subtle} domain shift. We note that the simplicity achieved ensures better explainability, and envision easier certification, as we avoid modifying the deep model in any way, but rather adapt an incoming image in a strictly defined fashion. 

A limitation of this study is the blunt selection of the \textit{k}-space low-frequency window, which could be improved by engaging intelligent search for the style-bearing spectrum components, such as presented in \cite{shipitsin2021gafl} for the supervised case, or by penalizing for the errors in the high-frequency part of the spectrum \cite{Pronina2020}. Another fundamental assumption we make is the separability of content and style, which is known to be true only partially \cite{ganin2016domainadversarial}. Optimization of the \textit{k}-space swapping pattern along with taking into account the intrinsic content-style coupling will be the subject of future work.

%% file: Sections/thanks.tex
\hfill Ivan Zakazov was supported by RSF grant 20-71-10134. Philips is the owner of the IP rights on the work described in this publication.

We warmly thank Prof.Kamnitsas for fruitful discussions in 2021 during early stages of this work.